\documentclass[twocolumn,aps,prb,widetext,showpacs,superscriptaddress]{revtex4}
\usepackage{mathrsfs,bm,multirow,amsmath,amsfonts,amssymb,array,booktabs,float}
\usepackage{graphicx,times,graphics,color,epsfig}
\begin{document}
\title{
Direct tunneling through high-$\kappa$ amorphous HfO$_2$: effects of chemical modification}

\author{Yin Wang}
\email{yinwang@hku.hk}
\affiliation{Department of Physics and the Center of Theoretical and Computational Physics, The University of Hong Kong, Pokfulam Road, Hong Kong SAR, China}
\author{Zhizhou Yu}
\affiliation{Department of Physics and the Center of Theoretical and Computational Physics, The University of Hong Kong, Pokfulam Road, Hong Kong SAR, China}
\author{Ferdows Zahid}
\affiliation{Department of Physics and the Center of Theoretical and Computational Physics, The University of Hong Kong, Pokfulam Road, Hong Kong SAR, China}
\author{Lei Liu}
\affiliation{Nanoacademic Technologies Inc., 7005 Blvd. Tachereau, Brossard, PQ, Canada, J4Z 1A7}
\author{Yu Zhu}
\affiliation{Nanoacademic Technologies Inc., 7005 Blvd. Tachereau, Brossard, PQ, Canada, J4Z 1A7}
\author{Jian Wang}
\affiliation{Department of Physics and the Center of Theoretical and Computational Physics, The University of Hong Kong, Pokfulam Road, Hong Kong SAR, China}
\author{Hong Guo}
\affiliation{Center for the Physics of Materials and Department of Physics, McGill University, Montreal, PQ, Canada, H3A 2T8}
\affiliation{Department of Physics and the Center of Theoretical and Computational Physics, The University of Hong Kong, Pokfulam Road, Hong Kong SAR, China}

\date{\today}
\begin{abstract}
We report first principles modeling of quantum tunneling through amorphous HfO$_2$ dielectric layer of metal-oxide-semiconductor (MOS) nanostructures in the form of n-Si/HfO$_2$/Al. In particular we predict that chemically modifying the amorphous HfO$_2$ barrier by doping N and Al atoms in the middle region - far from the two interfaces of the MOS structure, can reduce the gate-to-channel tunnel leakage by more than one order of magnitude. Several other types of modification are found to enhance tunneling or induce substantial band bending in the Si, both are not desired from leakage point of view.  By analyzing transmission coefficients and projected density of states, the microscopic physics of electron traversing the tunnel barrier with or without impurity atoms in the high-$\kappa$ dielectric is revealed.
\end{abstract}
\maketitle
\section{introduction}

Downscaling physical dimensions of metal oxide semiconductor (MOS) transistors has been the hallmark of electronics technology.\cite{ITRS, VLSI} As devices are now in the nano-meter scale, tunnel leakage through various barriers in MOS field-effect transistor (MOSFET) becomes a serious problem\cite{ITRS, VLSI, Frank} which led to the use of new barrier materials by industry, for instance the traditional dielectric material SiO$_2$ has been replaced by materials having higher permittivity (high-$\kappa$) in advanced technology. To maintain a necessary gate capacitance in MOSFET, the figure of interest is the ratio $\kappa/t$ where $\kappa$ is the permittivity and $t$ the physical thickness of the dielectric layer. In this sense a high-$\kappa$ dielectric having permittivity $\kappa_{h}$ and physical thickness $t_h$ is equivalent to a SiO$_2$ dielectric with permittivity $\kappa_{ox}$ ($\sim3.9$) and an \emph{equivalent oxide thickness} (EOT)\cite{Mistry} $t_{ox}=t_h\kappa_{ox}/\kappa_h$. Because $\kappa_h\gg\kappa_{ox}$, the physical thickness of the high-$\kappa$ dielectric $t_h$ can be made larger than that of SiO$_2$, therefore using high-$\kappa$ dielectric material can reduce tunnel leakage while maintaining the same gate capacitance. Another factor that affects tunnel leakage through the gate oxide is the barrier height since the top of valence band and bottom of conduction band of the barrier material determine the hole and electron tunneling, respectively. For SiO$_2$ the hole barrier is higher than the electron barrier, as a result the p-channel MOS (PMOS) has smaller gate leakage than the n-channel MOS (NMOS).\cite{Lo, Lee, Roy, Drazdziulis} For high-$\kappa$ dielectric HfO$_2$, the barrier height is 1.5 eV for electrons and 3.4eV for holes, to be compared to 3.1eV and 4.9eV of SiO$_2$.\cite{Cai}

Knowing the thickness and the potential height of an ideal smooth tunnel barrier, calculating electron transmission coefficient is a textbook problem of quantum mechanics.\cite{QM} For somewhat more complicated barriers, approximations such as the Wentzel-Kramers-Brillouin (WKB) approximation are widely applied, resulting to such well-known formula as the Fowler-Nordheim law.\cite{Lee, Lenzlinger, Yeo, QM} In these theories, the material, chemical and atomic details of the tunnel barrier are either ignored or parameterized into various phenomenological parameters. For atomically thin gate oxides, chemical details can play a very important role in the tunneling process\cite{Muller} and it is often unclear on how to parameterize microscopic details of the material in unique and/or physically meaningful manner. In this regard, atomic first principles calculations of tunneling is an important approach for understanding leakage through high-$\kappa$ dielectric barriers.

Atomistic calculations to extract tunneling physics are widely seen in the literature. In such analysis, one typically starts by calculating the tunneling potential using density functional theory (DFT),\cite{DFT} followed by calculating the tunnel conductance using a quantum transport theory such as linear response,\cite{LRT} scattering matrix,\cite{ST} scattering-state methods,\cite{SSM} the real-space finite-differencing,\cite{RSFDM} numerical solution of the Schr\"odinger equation,\cite{SE} and (nonequilibrium) Green's functions method.\cite{mGF,Nadimi,Chakraverty,Cartoixa,Liu}  Particularly, Refs.~\onlinecite{mGF,Nadimi,Chakraverty} reported calculations of tunneling in Si/B/Si structure where two bulk Si sandwiches a tunnel barrier (B);  Ref.~\onlinecite{Cartoixa} analyzed tunneling in M/B/M structures where M stands for metal. Analyzing the gate leakage in a nano-MOS necessarily requires analyzing structures having the form of Si/B/M, and Ref.\onlinecite{Liu} reports the first and perhaps so far the only first principles calculation of \emph{leakage current} through an \emph{amorphous barrier} between a metal and a semiconductor, i.e. the n-Si/SiO$_2$/Al MOS structure. Their results are well compared with those obtained from the traditional empirical formula - by using the potential parameters obtained from first principles calculation.\cite{Liu} In order to shorten the space charge region formed due to different work functions of Al and n-Si - so as to reduce the numerical cost, the Si in Ref.~\onlinecite{Liu} was extremely highly doped to $2\times10^{20}/cm^3$, significantly higher than the typical doping concentrations of MOSFETs and this underscores the difficulties of first principles device simulation.

In this work, we report first principles calculations of quantum tunneling through the technologically extremely important amorphous high-$\kappa$ dielectric HfO$_2$ in the prototypical Si/HfO$_2$/metal MOS structures. Our goal is to understand effects of microscopic chemical details to tunneling and predict possible chemical modifications that may reduce the gate-to-channel tunnel leakage.The effects to quantum tunneling by chemical modifications via atomic vacancies, impurities atoms and impurity positions are calculated from atomistic first principles. In particular we predict that chemically modifying the amorphous HfO$_2$ barrier by doping N and Al atoms in the middle - far from the two interfaces of the MOS structure, can reduce the gate-to-channel tunnel leakage by more than one order of magnitude. Several other types of modifications are found to enhance tunneling or induce substantial band bending in the Si, both are not desired from leakage point of view. By analyzing the transmission coefficients and projected density of states, we reveal the microscopic physics about electron traversing the tunnel barrier with or without impurity atoms in the junction.

The rest of the paper is organized as follows. In the next section the device model and calculation details are presented. Section III presents the results regarding effects of chemical modification to tunneling and Section IV is a short summary.

\section{Device model and structure relaxation}

To simulate gate-to-channel tunneling through a gate oxide in a MOS structure, in principle one may begin from a device structure where a gate oxide is sandwiched between a metal gate and a doped Si channel. By increasing (or decreasing) the gate voltage for NMOS (PMOS) to the threshold voltage so that the device changes from the accumulation state to the inversion state,\cite{VLSI, SM}, one calculates the gate-to-channel leakage. However, this ideal theoretical calculation is extremely difficult to realize by atomistic first principles methods, as the depletion layer or the space charge region in Si is so thick as to make calculations prohibitively expensive even with modern supercomputers. For instance, a highly but realistically doped Si with a doping concentration of $10^{18}/cm^3$ has a depletion layer around 50 nm thick.\cite{SM}  Much higher doped Si shortens the depletion layer thus reduces the computation cost as was done before.\cite{Liu} In addition, the work function difference between the metal and Si also contributes to the space charge in the Si/oxide interface, resulting in additional increase of the screening length in Si. Therefore, to investigate quantum tunneling in Si/HfO$_2$/metal structures from atomic first principle, we shall consider Si channels having reasonably high doping and consider an Al metal gate whose work function is reasonably close to that of the doped Si.

\begin{figure}
\includegraphics[width=\columnwidth]{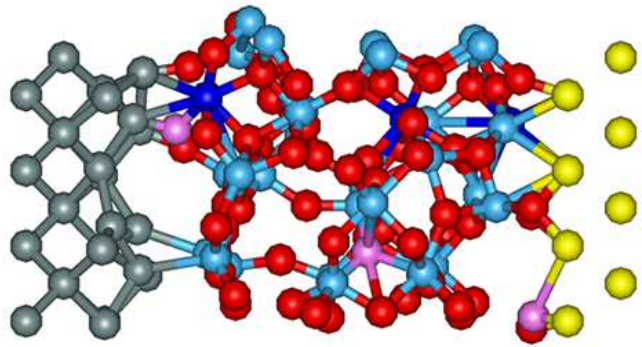}\\
\caption{(Colour online) The amorphous HfO$_2$ structure (red and pink spheres for O, light and dark blue spheres for Hf) sandwiched between Al (yellow) and n-Si (grey) leads. Chemical details of the amorphous HfO$_2$ are modified in some calculations. Pink (dark blue) spheres indicate one O (Hf) atom at the Si/HfO$_2$ interface (left interface), Al/HfO$_2$ interface (right interface), or inside the HfO$_2$ (middle region) are replaced by N (Al, Ta) or a vacancy.}
\label{fig1}
\end{figure}

Fig.~\ref{fig1} shows the atomic structure of the Si/HfO$_2$/Al MOS. Here, an amorphous HfO$_2$ is bonded between Si(100) on the left and Al(100) on the right (the $z$-direction) to form a tunneling structure. To analyze tunneling across the high-$\kappa$ barrier, we form a two-probe transport junction by extending the Si region to $z=-\infty$ and the Al region to $z=+\infty$, and make the structure periodic in the x-y directions.

To build the atomic model of Si/HfO$_2$/Al, an ensemble of ten samples of amorphous HfO$_2$ - each having 32 Hf atoms and 64 O atoms, are generated by a heat and quench method via classical molecular dynamics simulation as described in Ref.~\onlinecite{Wang}, and their lattice dielectric constants are calculated by DFT as implemented in the VASP package.\cite{Wang, VASP} Afterward we select the amorphous HfO$_2$ which has a homogeneous structure factor in all three directions and has lattice constants closest to the average value of the ensemble, as the barrier in the two probe Si/HfO$_2$/Al tunnel junction. The calculated lattice dielectric constant of this amorphous HfO$_2$ are 22.03, 27.30 and 26.52 in the x, y and z directions respectively, and the average value is close to the value that reported in Ref.~\onlinecite{Wang} and is consistent to the experimental value.\cite{Li} To form the scattering region of the two-probe Si/HfO$_2$/Al tunnel junction, four layers of Si and two layers of Al are connected to the HfO$_2$ as shown in Fig.~\ref{fig1}. The experimental lattice constant of 5.4307~\AA~is used for Si, and we strain the Al lattice ($\sim5.46\%$) to match the Si and HfO$_2$ region. The interfaces distances of Al/HfO$_2$ and Si/HfO$_2$ are optimized by VASP.\cite{VASP} By fixing all Al atoms and last two layers of Si atoms, the HfO$_2$ and two remaining layers of Si at the Si/HfO$_2$ interface are fully relaxed by DFT. Afterward, the relaxed Si/HfO$_2$/Al structure is extended from left and right by adding Si and Al crystal layers to form the scattering region of the two-probe system and, in total the whole scattering region in our transport calculations contains 800 atoms with amorphous HfO$_2$ ($\sim1.2 nm$) sandwiched between 80 layers of Si and 4 layers of Al, as shown in the inset of Fig.~\ref{fig2}. Note it is very important to have such a long Si to screen the potential of the scattering region when doing the transport calculations.

To investigate \emph{qualitative} effects of chemical details to the tunnel leakage, one oxygen (or Hf) atom in HfO$_2$ at the Si/HfO$_2$ interface (left interface), at the Al/HfO$_2$ interface (right interface), or in the middle of the HfO$_2$, were allowed to be replaced by a vacancy, or a nitrogen, an Al, or a Ta atom (Fig.~\ref{fig1}).  We shall denote such a replacement by notation A(B), namely Vac(Hf) stands for a vacancy replacing a Hf atom to produce a Hf vacancy; N(O) for one N atom to replace one O atom, etc.. To reduce computational cost we have ignored the local structure relaxation due to the replacement.

Having determined the atomic structure of the two-probe Si/HfO$_2$/Al tunnel junction, we calculate quantum transport properties by carrying out DFT within the nonequilibrium Green's function (NEGF) approach,\cite{NEGF} as implemented in the Nanodcal transport package.\cite{NEGFDFT,IEEEreview} For equilibrium calculations of the tunnel conductance, the NEGF reduces to the familiar Green's function approach. Standard norm-conserving nonlocal pseudopotentials\cite{PP} are used to define the atomic cores, a $s, p, d$ double-$\zeta$ plus polarization (DZP) linear combination of atomic orbital basis set\cite{DZP} is used to expand  physical quantities, and the exchange-correlation potential is treated at the local density approximation (LDA) level. For the two-probe NEGF-DFT self-consistent calculation of the density matrix and device Hamiltonian, a $4\times4$ k-mesh is applied to sample the 2-dimensional (2D, x-y) transverse Brillouin zone. Afterward, calculation of the transmission coefficient requires a much denser k-mesh, up to $100\times100$, in order to accurately determine this important physical quantity. In our transport calculations, the silicon lead is n-type doped to $5\times10^{18}/cm^3$ using the technique of virtual crystal approximation (VCA).\cite{VCA} This concentration is in the normal range of realistic MOS devices.\cite{ITRS}

In the rest of the paper we focus on analyzing the tunnel leakage in off-state PMOS (n-type Si) which is dominated by electron tunneling, and we qualitatively compare results among structures having different chemical details as presented above. On the other hand, for p-type Si lead both hole and electron contribute to tunnel leakage whose analysis requires accurate calculations of the band gap that is beyond the capability of the LDA functional used in this work. We note in passing that for the large number of atoms in the two-probe junction (e.g. 800), it remains a serious challenge and unsolved problem to apply higher level theory in first principles quantum transport simulation.

\section{Effects of chemical modification to tunneling}

\begin{figure}
\includegraphics[width=\columnwidth]{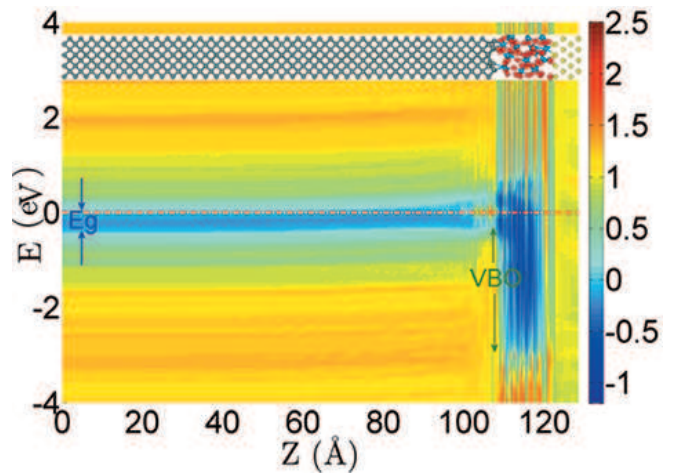}\\
\caption{(Colour online.) Projected density of states (PDOS) by different colours in logarithmic scale along the z-direction of n-Si/HfO$_2$/Al MOS structure. Red dashed line indicates the Fermi level, blue arrows show the band gap ($E_g$) of Si, and green arrows show the valence band offset (VBO) between Si and HfO$_2$. Top inset: atomic structure of the scattering region in the two-probe tunnel junction, grey atoms for Si, red atoms for O, light blue atoms for Hf and yellow atoms for Al.}
\label{fig2}
\end{figure}

Fig.~\ref{fig2} plots the calculated projected density of states (PDOS) along the z-direction of the Si/HfO$_2$/Al tunnel junction without any impurity replacement.\cite{PDOS} Several observations are in order. (i) The Fermi level locates at the bottom of the conduction band of silicon, consistent to the n-type doped material. (ii) The potential is well screened since the PDOS near the left and right boundaries of the scattering region are essentially constants. (iii) Band bending can be observed at the interface of n-Si and HfO$_2$ which is due to work function difference of the two materials. (iv) The valence band offset between silicon and HfO$_2$ is found to be about 2.6 eV, consistent with previous calculations.\cite{chen} The conduction band offset is small due to underestimation of the HfO2$_2$ band gap. In general, from Fig.~\ref{fig2} one can intuitively decern how a carrier (electron in our case) traverses from the n-Si lead to the Al lead by tunneling through the amorphous HfO$_2$ barrier.

\begin{figure}
\includegraphics[width=\columnwidth]{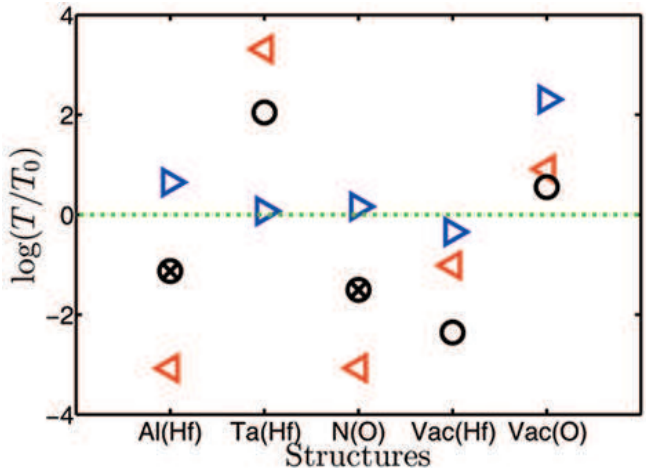}\\
\caption{(Colour online.) Transmission coefficient $T$ of the n-Si/HfO$_2$/Al tunnel junction having various chemical replacements normalized to that without impurity ($T=T_0$), plotted in logarithmic scale. Red left triangles, blue right triangles and black circles are transmission values of different structures with replaced atoms at left interface, right interface and middle HfO$_2$ region, respectively. Circles with a cross inside are the predicted structures that reduce tunnel leakage.}
\label{fig3}
\end{figure}

Fig.~\ref{fig3} shows the calculated transmission coefficients (at the Fermi level) for the n-Si/HfO$_2$/Al tunnel junctions with or without some impurity replacement. Recall we use the notation A(B) to denote B replaced by A. Here we observe that except for the Vac(O) case, replacing an atom at the Al/HfO$_2$ interface does not significantly affect transmission and, on the other hand, replacing an atom at the Si/HfO$_2$ interface or inside the amorphous HfO$_2$ affects transmission substantially. This is understandable because DOS at the Al/HfO$_2$ interface is very large due to the metal lead and transmission is therefore hardly affected by a sight change (due to the impurity) of charge density at this interface. Furthermore, oxygen vacancy usually introduces impurity levels\cite{Cartoixa, Umezawa} which may help increasing the transmission. In fact, as shown in Fig.~\ref{fig3} Vac(O) replacement is all situations increases transmission hence the tunnel leakage. Ta(Hf) impurity also increases tunneling while Al(Hf), N(O), and Vac(Hf) structures tend to decrease it.

\begin{figure}
\includegraphics[width=\columnwidth]{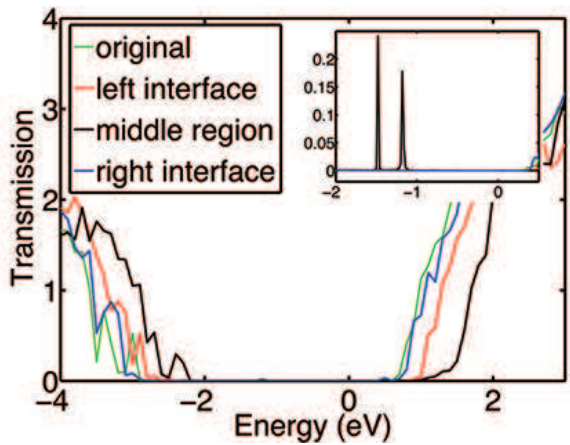}\\
\caption{(Colour online.) Transmission versus energy curves with the energy interval of 0.1 eV for the n-Si/HfO$_2$/Al  junction without impurity (green), and for structures having the N(O) replacement at the left interface (red), right interface (blue), or the middle region (black). Inset: Transmission versus energy curves (energy interval of 0.02 eV) inside the band gap for the n-Si/HfO$_2$/Al structures. Only N(O) structure in the middle region has impurity energy levels (black) inside the gap, which contributes to tunneling.}
\label{fig4}
\end{figure}

\begin{figure}
\includegraphics[width=\columnwidth]{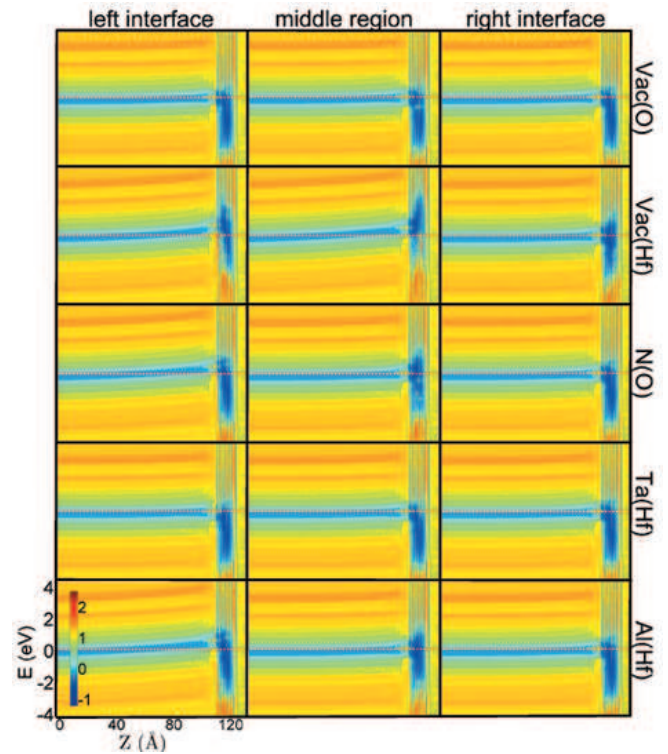}\\
\caption{(Colour online.) Projected density of states along the z-direction of all the structures having an impurity atom or vacancy replacement. Red dashed line indicates the Fermi level, all the figures have the same axes as the lower left figure for the Al(Hf) structure at the left interface. Color coding values are given by the vertical bar in the lower left figure.}
\label{fig5}
\end{figure}

To better understand how microscopic chemical details affect tunneling, we plot the calculated transmission coefficient versus energy $T=T(E)$. Fig.~\ref{fig4} shows the $T(E)$ curves with the energy interval of 0.1 eV for a tunnel junction without impurity replacement (green) and one with N(O) replacement at three regions. We observe that the electron barrier of the original structure (e.g. no impurity replacement) is the lowest compared to others, followed by N(O) structure at the right interface, left interface and the middle  region. The tunneling is therefore largest for the original structure than those with the N(O) replacement, as seen earlier in Fig.~\ref{fig3}. It is interesting to observe that N(O) replacement in the middle region gives the largest barrier but not the smallest transmission, because of the impurity level inside the band gap contributes to tunneling (see inset of Fig.~\ref{fig4} with the energy interval of 0.02 eV). Examining all the other tunnel junctions we conclude that at the microscopic level, chemical details by impurity replacement alters the scattering potential and induces impurity level, both contribute to the tunneling.

PDOS for several tunnel junctions are plotted in Fig.~\ref{fig5} which well explain the magnitude of tunneling. Compared with the junction without impurity replacement, the Ta(Hf) and Vac(O) replacements give larger tunneling by reducing the potential barrier. Vac(Hf) and Al(Hf) replacements at left interface and in the middle region give higher tunnel barrier, hence these structures have smaller tunneling. Note that a band bending induced inversion layer of Si at the Si/HfO$_2$ interface can be clearly seen from the PDOS of the N(O) replacement at left, Vac(Hf) replacements at left and middle, as well as Al(Hf) at left, which results to a thicker barrier in the Si region due to the inversion of Si (see Fig.\ref{fig5}). While this is desirable for reducing the gate-to-channel tunnel leakage, the inversion of Si at the interface of Si/HfO$_2$ will increase the source-to-drain leakage in the off-state. Therefore and on balance, these structures are not the most desired. We therefore conclude that among all the structures we have investigated, the N(O) and Al(Hf) replacements in the middle region are promising structures to reduce the gate-to-channel tunnel leakage.

\section{summary}

In this work we have investigated the physics of tunnel leakage in MOS structures made of the technologically important amorphous high-$\kappa$ dielectric, from atomistic first principles without any phenomenological parameter. To the best of our knowledge, this is the first time such an investigation is carried out. Our goal is to understand effects of microscopic chemical details to tunneling and predict possible chemical modifications that may reduce the gate-to-channel tunnel leakage. We found that atomic impurities - when doped at proper regions in the MOS structure, can significantly reduce tunneling by more than an order of magnitude. In particular we predict that chemically modifying the amorphous HfO$_2$ barrier by doping N and Al atoms in the middle - far from the two interfaces, can reduce the gate-to-channel tunnel leakage substantially. By analyzing the transmission coefficients and projected density of states, we reveal the microscopic physics about electron traversing the tunnel barrier with or without impurity atoms in the junction. Finally, we mention in passing that it is important to analyze structures having a semiconductor and a metal contacts in order to understand the tunnel leakage in MOS structures, namely systems such as the n-Si/HfO$_2$/Al studied here. This is because the band bending in the Si - induced by work function difference of the MOS layers and also by impurity doping in the barrier region, crucially affects the tunneling as we have found.

{\bf Acknowledgements.} YW are grateful to Dr. Jianing Zhuang and Mr. Ronggen Cao for useful discussions concerning the physics of band bending. This work is supported by the University Grant Council (Contract No. AoE/P-04/08) of the Government of HKSAR, NSERC (HG), IRAP (YZ, LL) of Canada. We thank CLUMEQ and Compute-Canada for the computation resources.


\begin{thebibliography}{100}
\bibitem{PDOS}
The PDOS are first projected on each atom, and then plotted along the MOS leakage transport direction (z direction).
\end{thebibliography}
\end{document}